\documentclass[aps,pra,twocolumn,epsfig,amsfonts,amsmath,amstex,floatfix]{revtex4}

\usepackage{lipsum}
\usepackage{bbold}
\usepackage{color}

\usepackage{amsfonts,amsmath}
\usepackage{epsfig,amsmath}
\usepackage{graphicx}
\usepackage{dcolumn}
\usepackage{bm}
\usepackage{bbold}
\newcommand{\beq}{\begin{equation}}
\newcommand{\eeq}{\end{equation}}
\newcommand{\beqa}{\begin{eqnarray}}
\newcommand{\eeqa}{\end{eqnarray}}

\newcommand{\la}{\langle} 
\newcommand{\ra}{\rangle}

\def\jpb#1{{ J.\ Phys.\ B} {\bf#1}}
\def\jpa#1{{ J.\ Phys.\ A} {\bf#1}}

\def\nat#1{{ Nature} {\bf#1}}

\def\njp#1{{ New\ J.\ Phys.} {\bf#1}}

\def\pr#1{{ Phys.\ Rev. } {\bf#1}}
\def\pra#1{{ Phys.\ Rev. A\/} {\bf#1}}
\def\prb#1{{ Phys.\ Rev. B\/} {\bf#1}}

\def\prl#1{{ Phys.\ Rev.\ Lett.} {\bf#1}}

\def\rmp#1{{ Rev. \ Mod. \ Phys.} {\bf#1}}

\begin{document}

\title{Freezing and Thawing of Entanglement in Lossless Multiparty Systems}
\author{X.-F. Qian $^{1}$}
\email{xqian6@stevens.edu}
\author{C. Qu $^{1}$}
\author{J.H. Eberly $^{2}$}
\affiliation{$^{1}$Center for Quantum Science and Engineering, and Department of Physics, Stevens Institute of Technology, Hoboken, New Jersey 07030, USA\\
$^{2}$Center for Coherence and Quantum Optics, and Department of Physics \& Astronomy, University of Rochester, Rochester, NY 14627, USA}

\date{\today}

\begin{abstract}
Entanglement freezing has been demonstrated existing in various noisy decoherence mechanisms. Here we explore its universality by investigating freezing behavior in a lossless multiparty system, i.e., an $N$-site optical lattice (or equivalently spin-chain) system. Interesting phenomenon of permanent entanglement freezing is found for the infinite-size case ($N\rightarrow\infty$). As the multiparty system size reduces to finite $N$, the interesting behavior of dynamical entanglement ``thawing" start to emerge. Especifically, alternative appearances of ``freezing" and ``thawing" can be identified as time evolves. Our results may provide useful guidance to entanglement control in quantum tasks. 
\end{abstract}



\maketitle

\noindent {\bf Introduction:} Correlations can be said to be the backbone of mathematical physics. Quantum entanglement is a special form of multi-party correlation,  known for decades as a powerful non-classical resource, essential for quantum events such as teleportation and for inclusion in processes designed to assist quantum communication and quantum computing \cite{NC2000, Horodecki2009RMP}. Coherence itself, as a correlation feature definable in any linear vector space,  has also been identified as a resource (see \cite{Streltsov2017RMP}).

A newly discovered measure of correlation  \cite{OlZur, HendVedr}  named quantum discord is now  regarded as a generalization of entanglement in the following surprising way. Some mixed states that have nonzero discord, previously considered as classical states because entanglement is absent, have nevertheless been reported to be available for non-classical computational speed-up \cite{Datta}.

Equally unusually, for certain quantum systems, even if entanglement has already decayed to zero, their quantum discord can persist for a long time before decaying  suddenly \cite{Mazzo}. This temporary steady persistence is labeled as the ``freezing" of quantum discord. It has been studied together with ``thawing", the equally sudden onset of discord's decay (see \cite{Maziero2009PRA, Pinto, Cianciar}).  Experimental observations confirm the existence of these sudden transitions between freezing and thawing \cite{Silva, Paula}. Further study has shown that entanglement and coherence can also be subject to freezing  \cite{AliRau, Bromley2015PRL, WuXu, Ding}.

All of these studies have been concerned with freezing in the active presence of one or another decoherence mechanism. However, in most quantum tasks information is often processed in finite-size multiparty network systems \cite{NC2000, Kimble2008Nature}. A natural question arises: is freezing a universal phenomenon for all size systems? In this report we explore entanglement freezing in a lossless $N$-site optical lattice system. Besides freezing, the interesting complementary phenomenon of entanglement ``thawing" is also identified. In addition, the size of the system (determined by $N$) is playing a crucial role in controlling proportionally the length of the freezing time interval. Our result refutes the widespread impression that entanglement freezing and thawing occur as a consequence of an available decoherence, and illustrates their universality in lossless multiparty systems. 

For the convenience of discussion, we define a ``freezing" process as a ``stable" behavior within a time interval where the physical quantity such as entanglement $E(t)$ is approximately independent of time with $\partial E(t)/\partial t \approx 0$ or the fluctuation is relatively negligible comparing to the (relative) value of the quantity, i.e., $\sigma (E)/\Delta E \ll 1$.  \\

\noindent {\bf Multiparty optical lattice context:} Now we describe a multiparty context where the track of entanglement information flow is under consideration. We study a scenario where a system of interest (a qubit $A$) is interacting with a multiparty network (an optical lattice chain $B$). The most general initial state (at the time of interest $t=0$) of a qubit is an arbitrary mixed state, i.e., 
\begin{equation}
\rho _{A}=\left(
\begin{array}{cc}
\rho _{ee} & \rho _{eg} \\
\rho _{ge} & \rho _{gg}
\end{array}
\right).  \label{Mixed}
\end{equation}
Here the qubit is described in its excited and ground state basis, $|e\ra$, $|g\ra$. We ascribe without loss of generality that this mixedness is caused by interaction of previously interacting partners during a preparation process that may include the physical qubit host, experimental device, previous environment, etc., which we lump together and label as the marginal party $M$. Then this unspecified marginal $M$ naturally purifies the system state, i.e., $|\psi _{AM}\rangle =\cos\theta|e\rangle \otimes |m_1\rangle + \sin\theta |g\rangle \otimes |m_2\rangle$, where $\la m_1|m_1\ra=\la m_2|m_2\ra=1$ describe normalized marginal states. Here $\cos\theta=\sqrt{\rho _{ee}}$, $\sin\theta=\sqrt{\rho _{gg}}$ and $\la m_1|m_2\ra=\alpha=\rho_{eg}/\sqrt{\rho _{ee}\rho _{gg}}$ charactering a generic overlap determined by the arbitrary density matrix (\ref{Mixed}).

After the qubit system is prepared, it is then sent to participate in certain tasks run by a quantum network partner $B$, which can be reasonably treated as separable (due to no previous interaction) with the system. Thus the overall initial state can be described as
\begin{equation}
|\psi _{ABM}(0)\rangle =(\cos\theta|e\rangle \otimes |m_1\rangle + \sin\theta |g\rangle \otimes |m_2\rangle)|\phi_0\ra, \label{initial}
\end{equation}
where $|\phi_0\ra$ is the initial quantum net work state. 

The total Hamiltonian is described as 
\begin{equation}
H=H_{A}+H_{B}+H_{I}+H_{M},
\end{equation}
where $H_{A}$, $H_{B}$ and $H_{M}$ are the Hamiltonians of qubit system $A$, quantum network $B$ and marginal $M$ respectively. $H_{I}$ denotes the currently active interaction after $t=0$, between qubit $A$ and its quantum network partner $B$. Here we consider the multiparty network system $B$ as $N$ ultracold atoms confined in an optical lattice formed by several standing-wave laser beams. In the nearest neighbor consideration, such a general Hamiltonian (see Ref.~\cite{Duan2003PRL}) can be written as
\beqa
H_{B}&=&-\sum_{\langle i j\rangle s}\left(t_{\mu s} a_{i s}^{\dagger} a_{j s} +{\rm H.c.} \right) \notag \\
&+ &\frac{1}{2} \sum_{i, s} U_{s} n_{i s}\left(n_{i s}-1\right) + U_{\uparrow \downarrow} \sum_{i} n_{i \uparrow} n_{i \downarrow}, \label{cold-atom}
\eeqa
where the summation index $\langle i j\rangle$ indicates nearest neighbor interactions, $a_{i s}$ are annihilation operators, $n_{i s}=a_{i s}^{\dagger} a_{i s}$ are number operators, and the spin parameter $s=\uparrow, \downarrow$ denotes two relevant internal states of each atom. Here $t_{\mu s}$ and $U_{s}$ ($U_{\uparrow \downarrow}$) are tunneling and on-site interaction energies depending on laser-created potentials, spin orientations, and lattice cubic geometry directions $\mu=x,y,z$. In the controlled parameter regime $t_{\mu s} \ll U_{s}, U_{\uparrow \downarrow}$ and $\la n_{i\uparrow }\ra +\la n_{i\downarrow }\ra \approx 1$ (representing the single-occupation insulating phase), the above Hamiltonian (\ref{cold-atom}) is equivalent to an antiferromagnetic or ferromagnetic Heisenberg spin XXZ model \cite{Duan2003PRL}. 

As an illustration, we consider a chain type of network, i.e, all atoms are aligned in a chain with each atom interacting only with its two nearest neighbors. Then the ultracold atom at one end of the chain is treated as qubit $A$ and the remainder of the lattice-trapped atoms as the interacting network $B$. Such a scenario can be understood as a prepared atom $A$ in site 1 that is placed to initiate a quantum task with the assistance of the interacting lattice chain $B$. We take the case $U_{s}=2 U_{\uparrow \downarrow}$ and keep the leading order in $t_{\mu s}/U_{\uparrow \downarrow}$ through the Schrieffer-Wolff transformation \cite{Schrieffer-Wolff} (see also in \cite{Duan2003PRL, Hewson1997}). Then the Hamiltonian $H_{AB}=H_A + H_B + H_{I} $ can be described as 
\begin{eqnarray}
H_{AB} &=&\eta_{\mu \perp}(\sigma _{A}^{+}\sigma _{1}^{-}+\sigma
_{A}^{-}\sigma_{1}^{+})  \notag \\
&+&\sum_{i=1}^{N} \eta_{\mu \perp} (\sigma _{i}^{+}\sigma _{i+1}^{-}+\sigma
_{i}^{-}\sigma _{i+1}^{+}). \label{XY}
\end{eqnarray}
Here $\sigma ^{\pm }=\sigma ^{x }\pm i\sigma ^{y} $ are usual raising and lowering Pauli
matrices with  $\sigma ^{x }=a_{\uparrow}^{\dagger} a_{\downarrow}+a_{\downarrow}^{\dagger} a_{\uparrow} $, $\sigma ^{y}=-i(a_{\uparrow}^{\dagger} a_{\downarrow}- a_{\downarrow}^{\dagger} a_{\uparrow} )$, and $\eta_{\mu \perp}=t_{\mu \uparrow}t_{\mu \downarrow}/2U_{\uparrow \downarrow}$. The structure of such an ultracold atom optical lattice model (\ref{XY}) is equivalent to that of an XY spin model. 

We now assume that all the $N$ lattice sites in interaction partner $B$ are initially in their lower energy state, i.e., in the eigenstates (which we denote as $|0_i\ra$) of the number operators $n_{i \downarrow}$. Then the overall system $B$ initial state is given as $|\phi_0\ra=|0_1\ra...|0_i\ra...|0_N\ra\equiv|{\bf 0}\ra$. 

The eigenstates and eigenenergies of the above Hamiltonian (\ref{XY}) can be obtained as \cite{Qian2005PRA}
\beqa
|k\rangle &=&\sqrt{\frac{2}{N+2}}\sum_{j=1}^{N}\sin \left( \frac{jk\pi }{N+2}
\right) |{\bf 1}_{j}\rangle, \\
E_{k}&=&2\eta_{\mu \perp}\cos \left( \frac{k\pi }{N+2}\right),
\eeqa
where $k\in \left[ 1,N\right]$ and \{$|{\bf 1}_{j}\ra$\} are the states of the multi-site system $B$ indicating the $j$-th atom is in the spin up state while
all the remaining ones are spin down. 

Then one can straightforwardly obtain the time dependent state through the unitary operator $U(t) =\sum_{k=1}^{N}e^{-iE_{k}t}|k\ra\la k|$ from the initial state (\ref{initial}), i.e.,
\begin{eqnarray}
|\psi _{ABM}(t)\ra  &=&\cos\theta |m_{1}(t)\ra \Big(c_{e}(t)|{\bf 0}\ra
|e \ra + \sum_{j=1}^{N} c_{j}(t) |{\bf 1}_{i}\ra |g\ra
\Big)
\notag \\
&+& \sin \theta |m_{2}(t)\ra  |{\bf 0}\ra |g \ra, \label{time}
\end{eqnarray}
where we have defined the time dependent coefficients
\beqa \label{ce}
c_{e}(t)&=&\sum_{k=1}^{N+1}\frac{2e^{-iE_{k}t}}{N+2}\sin \left(
\frac{k\pi }{N+2}\right) \sin \left( \frac{k\pi }{N+2}\right), \notag\\
c_{j}(t)&=&\sum_{k=1}^{N+1}\frac{2e^{-iE_{k}t}}{N+2}\sin \left(
\frac{k\pi }{N+2}\right) \sin \left( \frac{j+1}{N+2}k\pi \right) \label{cn}
\eeqa
with $c_{e}(0)=1$, $c_{j}(0)=0$ and $j=1,2,..., N$.\\


\noindent {\bf Entanglement freezing and thawing:} Since atom $A$ is the key initiation qubit and the multi-site optical lattice chain $B$ is an integrated assisting system, it is natural to consider entanglements focused on these two parties respectively. Based on the time-dependent multiparty state (\ref{time}), we analyze two dynamical entanglements, i.e., $K_A(t)$ between qubit $A$ and the $BM$ remainder, and $K_B(t)$ between the optical lattice chain $B$ and the $AM$ remainder. The multi-dimensional pure state nature of state (\ref{time}) suggests the optimal measure to adopt is Schmidt weight \cite{Grobe, Eberly2006}, which is defined as
\beq
K=\frac{1}{\sum_j \lambda^2_{j}},
\eeq
where $\lambda_j$ are the eigenvalues of the reduced density matrix $\rho_Q$ with $Q=A,B$ by tracing out the remaining parties. 

The reduced density matrix of the system $A$ can be obtained as  $\rho_A={\rm Tr}_{BM} |\psi_{ABM}(t)\ra\la\psi_{ABM}(t)|$, i.e., 
\begin{equation}
\rho _{A}=\left(
\begin{array}{cc}
\cos^2\theta |c_e(t)|^2 & \sin\theta\cos\theta c_e(t)\alpha^*\\
\sin\theta\cos\theta c^*_e(t)\alpha & \sin^2\theta+\cos^2\theta \sum_{i=1}^{N} |c_{i}(t)|^2
\end{array}
\right). 
\end{equation}
Then one can achieve the dynamical entanglement as
\beqa
K_A(t)=\frac{2}{1+[1-f(t)]^2 +2\sin^2\theta f(t)|\alpha|^2 }, \label{KA}
\eeqa
where $f(t)=2\cos^2\theta |c_e(t)|^2$. 

Similarly, one can achieve the reduced density matrix for party $B$ and consequently compute the Schmidt weight $K_B(t)$ as 
\beqa
K_B(t)=\frac{2}{1+(1-g(t))^2 +2\sin^2\theta g(t)|\alpha|^2 }, \label{KB}
\eeqa
where  $g(t)=2\cos^2\theta \sum_{j=1}^{N} |c_{j}(t)|^2$. 

\begin{figure}[h]
\includegraphics[width=6.2 cm]{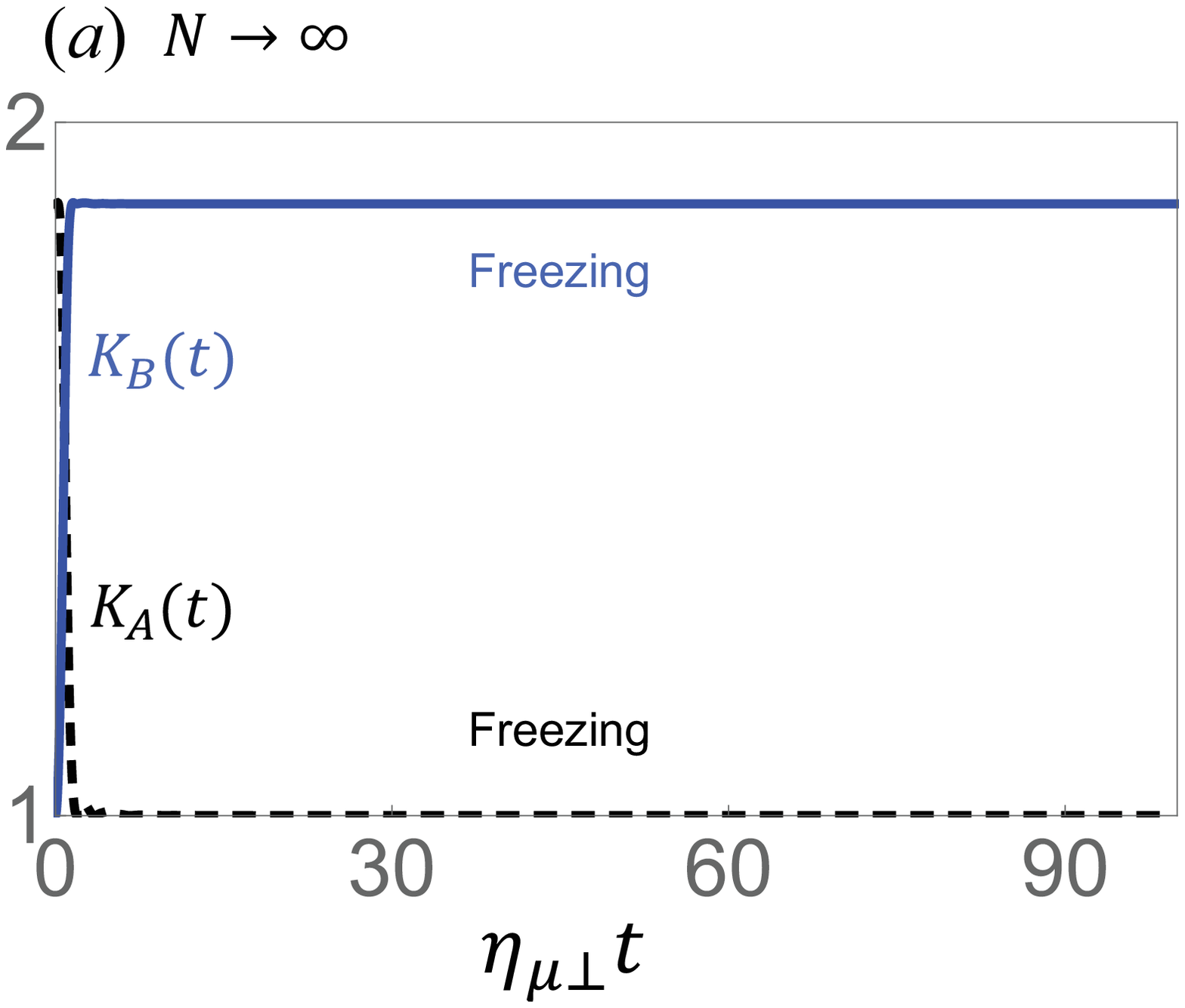}
\includegraphics[width=6.2cm]{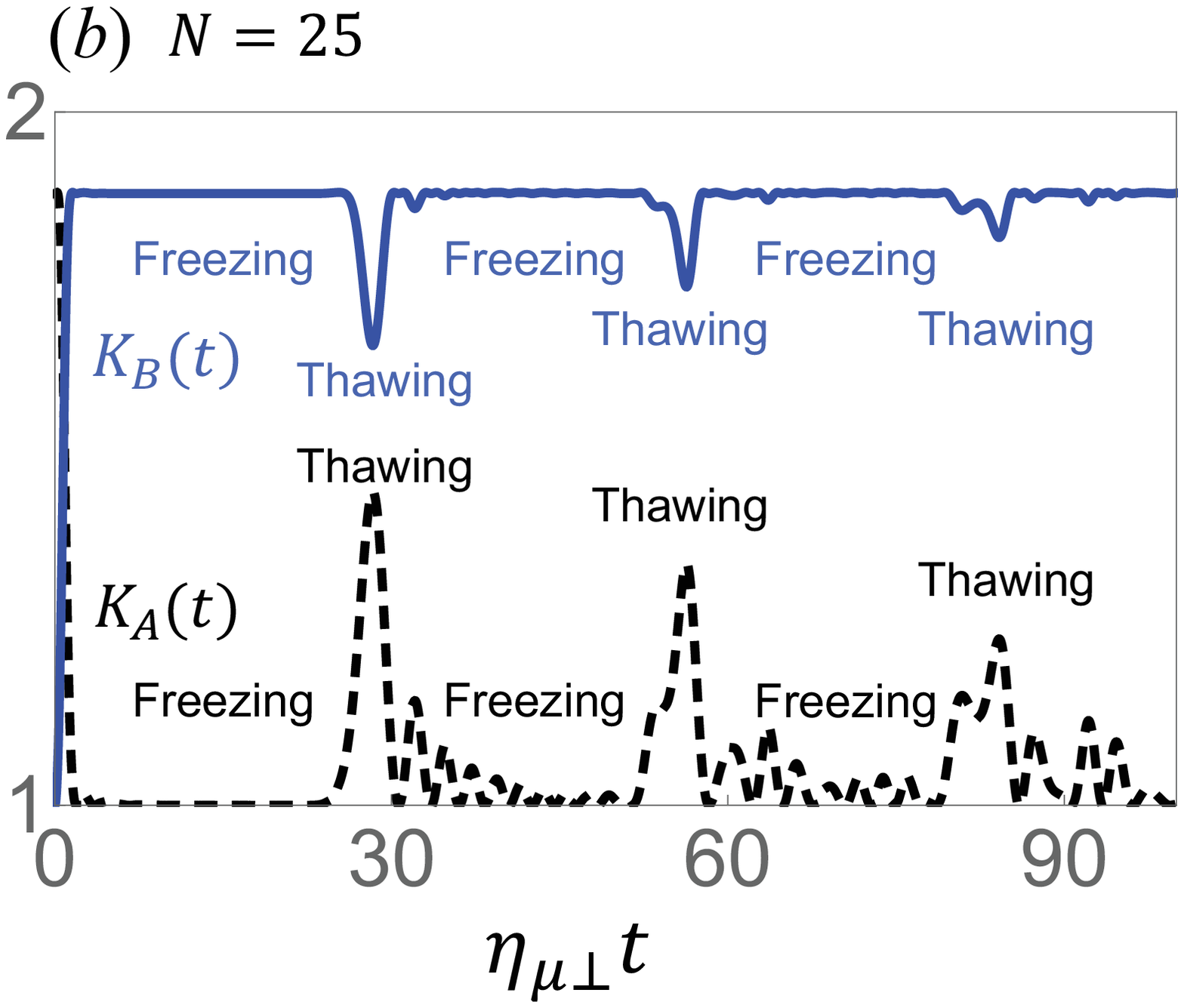}
\includegraphics[width=6.2cm]{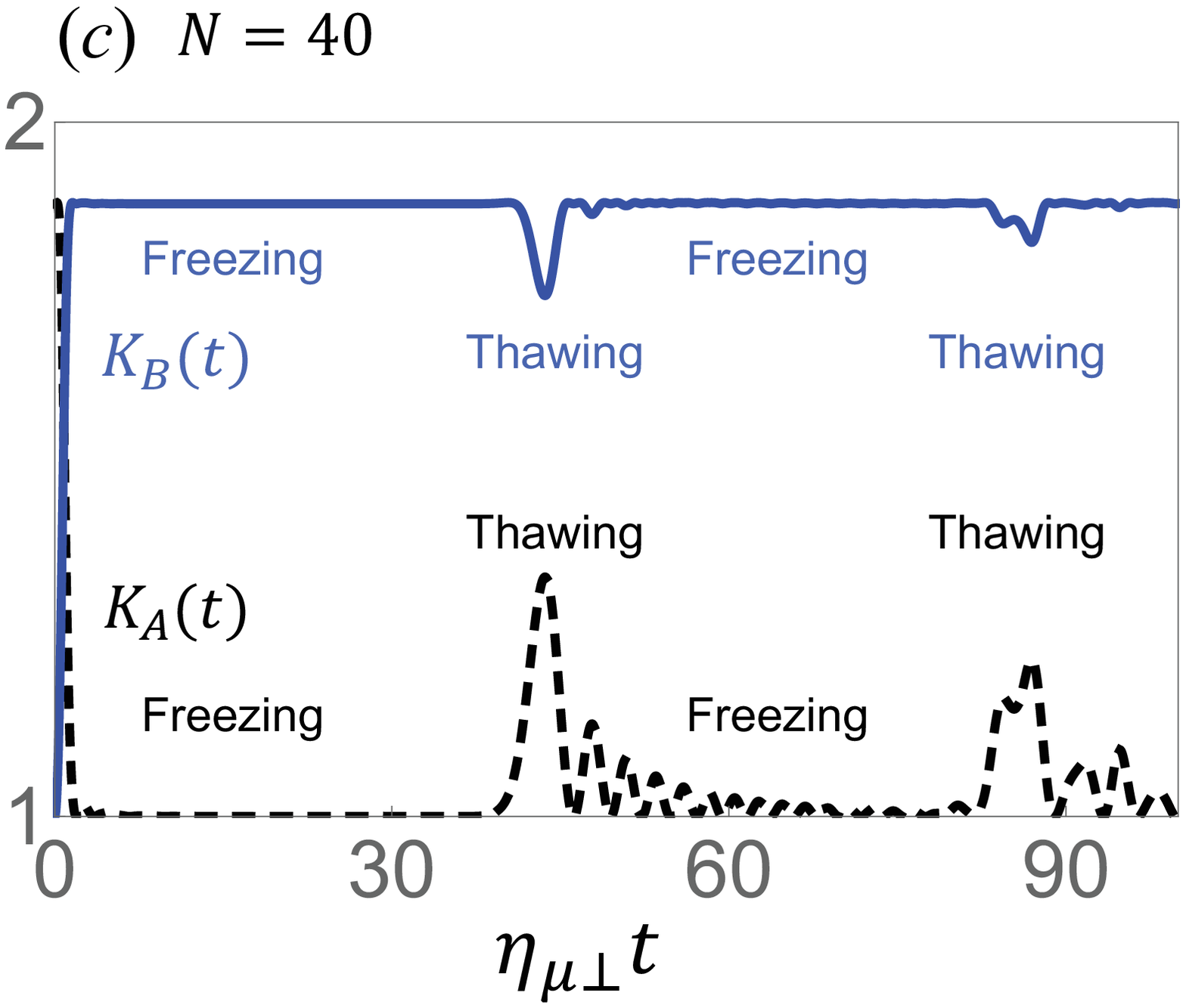}
\caption{Entanglement freezing and thawing behaviors of Schmidt weights $K_{A}(t)$ and
$K_{B}(t)$. Panels (a), (b), and (c) illustrate entanglement dynamics for $N\rightarrow \infty$, $N=25$, and $N=40$ respectively. The dashed black and solid blue lines denote correspondingly $K_{A}(t)$ and $K_{B}(t)$, and the $x$ axis is in units of $\eta_{\mu \perp}t$. Permanent freezing is phenomena is identified for both entanglements when $N\rightarrow \infty$. Entanglement thawing/defreezing occurs in the finite $N$ cases at approximately periodically at times $\eta_{\mu \perp}t=k(N+2)$ with arbitrary integer $k$. In all cases, the first freezing period starts at approximately $\eta_{\mu \perp}t=2$.}
\label{freezing}
\end{figure}

In both cases, there are two effective number of eigenvalues of the reduced density matrices. Thus both $K_A(t)$ and $K_B(t)$ take values between 1 and 2, where 1 means no entanglement and 2 stands for maximum entanglement \cite{Qian2018NJP}. 

To analyze entanglement freezing behavior of our lossless multiparty scenario in comparison with the noisy decoherence mechanism, we first consider the closest case of the two, i.e., when the lattice chain system contains infinite number of atoms with $N\rightarrow \infty $. Interestingly, one can show in this limiting case, the time-dependent amplitude reduces to a compact form
\beq
c_{e}(t)=\frac{J_{1}(2\eta_{\mu \perp}t )}{\eta_{\mu \perp}t},
\eeq
where $J_{1}(. )$ is the standard Bessel functions of the first kind. The detailed proof of the this relation is given in Appendix A, see also a similar analysis in \cite{PrattEberly}.

Then the two dynamical entanglements $K_A(t)$, $K_B(t)$, in this limiting case $N\rightarrow \infty$, are simple functions of the Bessel function $J_{1}(2\eta_{\mu \perp}t )$. Their behaviors with respect to the time unit $\eta_{\mu \perp}t$ are illustrated in Fig.~\ref{freezing} (a). Permanent freezing behavior can be identified for both entanglements, which is similar to the results obtained with noisy decoherence mechanisms in previous studies \cite{AliRau, Bromley2015PRL, WuXu}. To prove analytically the freezing behavior, one will need to show the vanishing of the derivatives, i.e., $\frac{\partial K_A(t)}{\partial t}\approx 0$ and $\frac{\partial K_B(t)}{\partial t}\approx 0$ which is done in detail in Appendix. B.

We now further analyze the effect of the system size (i.e., the finite number of ultracold atoms $N$ in the multi-site optical lattice chain $B$) on entanglement freezing. The specific behaviors of both $K_A(t)$ and $K_B(t)$ are illustrated in detail in Fig.~\ref{freezing} for two cases $N=25, 40$. The permanent freezing behavior in (a) disappears and is replaced with alternative occurrences of both freezing and thawing. The first freezing and first thawing occur approximately at 
\beq
t_{freeze}=\frac{2}{\eta_{\mu \perp}} \quad {\rm and} \quad t_{thaw}=\frac{N+4}{\eta_{\mu \perp}} \label{timing}
\eeq 
respectively. The analytical derivation of the freezing and thawing timings are given in detail in Appendix C. 

Surprisingly, from comparing the three plots in Fig.~\ref{freezing}, one notes that the dynamical entanglements $K_A(t)$, $K_B(t)$ for finite $N$ are almost identical to those of the infinite $N$ case before the first thawing happens at $\eta_{\mu \perp}t=N+4$. This can be confirmed by comparing the significant terms in the Taylor expansion of the time-dependent amplitude $c_{e}(t)$ with respect to time $t$ for both finite and infinite $N$ (see detailed analysis in Appendix D). Such a behavior demonstrates directly that entanglement freezing is preserved for multiparty lossless systems, and it is not a unique phenomenon of noisy decoherence mechanisms. The first occurrence of freezing is independent of the size of the multiparty system but the length of the freezing time interval is however strongly controlled by the size $N$ as shown in (\ref{timing}).

Furthermore, the alternative occurrence of freezing and thawing has an approximate period,
\beq
T=\frac{N+2}{\eta_{\mu \perp}},
\eeq
which is proportional to the size $N$ of the system. A detailed derivation of this period is also given in Appendix C. From this result, one notes entanglement thawing happens when $\eta_{\mu \perp}t=k(N+2)+2$ with arbitrary integer $k$. These properties are demonstrated in the dynamical plots in Fig.~\ref{freezing}.\\




\noindent {\bf Summary and discussion:} In conclusion, we have identified the persistence of entanglement freezing dynamics in a lossless multiparty system hat consists of $N$-site ultracold atoms in an optical lattice chain. This indicates the universality of this entanglement dynamical phenomenon in both lossy and lossless contexts. When the size of the multiparty system is infinitely large, permanent freezing behavior is retrieved just as in lossy decoherence scenarios. Although the freezing phenomenon exists for finite size (finite $N$) lossless contexts, its behavior has to be balanced by the alternative occurrence of entanglement thawing or defreezing. 

Entanglement freezing in lossless multiparty system can be very useful for quantum information tasks, most of which require a relatively stable behavior of entanglement. This will largely release the strict restrictions, thus providing huge margins, on the timing of information processing and computation. The obtained specific size-dependent freezing and thawing time parameters, $t_{freeze}, t_{thaw}, T$ may provide useful guidance to quantum control. 

We expect entanglement freezing and thawing dynamics to be found in various other physical contexts. \\

\noindent {\bf Acknowledgement:} We acknowledge discussions with Y. Ding and S.B Xie, as well as financial support from NSF grants PHY-1505189, INSPIRE PHY-1539859, and Stevens Institute of Technology.


\section{Appendix}
\subsection{Bessel function analysis}
Here we now analyze the behavior of the time dependent coefficient $c_e(t)$ given in equation (\ref{ce}) in the limiting case when $N\rightarrow \infty$. At the large $N$ case, one can transfer the summation into an integral by defining
\begin{equation}
\frac{1}{N+2}=dy\text{, \ }\frac{k}{N+2}=y,
\end{equation}
which leads to the integration expression
\begin{eqnarray}
c_{e}\left( t\right)  &=&2\int_{\frac{1}{N+2}}^{\frac{N+1}{N+2}}dy\exp \left[
-i2\eta_{\mu \perp}t\cos \left( y\pi \right) \right] \sin ^{2}\left( y\pi \right) \notag \\
&=&-\frac{2}{\pi }\int_{0}^{1}\exp \left[ -i2\eta_{\mu \perp}t \cos \left( y\pi \right) 
\right] \sin \left( y\pi \right) d\cos \left( y\pi \right)  \notag \\
&=&-\frac{2}{\pi }\int_{1}^{-1}\exp \left[ -i2\eta_{\mu \perp}t x\right] \sqrt{1-x^{2}} dx \notag \\
&=&\frac{1}{\eta_{\mu \perp}t }J_{1}(2\eta_{\mu \perp}t)
\end{eqnarray}
where $x=\cos \left( y\pi \right)$. 

\subsection{Entanglement freezing for infinite N}
In the following we compute the derivative of entanglement $K_B(t)$ respect to time for the multi-site lattice chain system $B$. From the time dependent entanglement (\ref{KB}), one has  
\beqa
\frac{\partial K_B(t)}{\partial t}&=&\frac{-2}{[1+(1-g(t))^2 +2\sin^2\theta g(t)|\alpha|^2]^2 } \notag\\
&&\frac{\partial [1+(1-g(t))^2 +2\sin^2\theta g(t)|\alpha|^2]}{\partial t} \notag\\
&=&\frac{-2[-(1-g(t))g'(t)+2\sin^2\theta g'(t)|\alpha|^2]}{[1+(1-g(t))^2 +2\sin^2\theta g(t)|\alpha|^2]^2 } \notag\\
&=&\frac{2[(1-g(t))-2\sin^2\theta |\alpha|^2][-f'(t)]}{[1+(1-g(t))^2 +2\sin^2\theta g(t)|\alpha|^2]^2 } \notag\\
&=&\frac{8[2\sin^2\theta |\alpha|^2-1+g(t)][\cos^2\theta c'_e(t)]}{[1+(1-g(t))^2 +2\sin^2\theta g(t)|\alpha|^2]^2 } , \label{derivativeKB}
\eeqa
where we have used the fact $g'(t)=-f'(t)$ because $g(t)=2\cos^2\theta |c_g(t)|^2=2\cos^2\theta(1- |c_e(t)|^2)=2\cos^2\theta-f(t)$.

From the properties of Bessel function derivatives, i.e., 
\beq
\frac{\partial J_\nu (x)}{\partial x} =\frac{\nu}{x}J_\nu (x)-J_{\nu+1} (x),
\eeq
where $\nu$ is the order of the Bessel function, one can straightforwardly obtain the following derivative
\beq
\frac{\partial c_e(t)}{\partial t}=\frac{\partial }{\partial t}\frac{J_{1}(2\eta_{\mu \perp}t )}{\eta_{\mu \perp}t}=-\frac{J_{2} (\eta_{\mu \perp}t)}{\eta_{\mu \perp}t}\approx 0.
\eeq
When combined with the result obtained in (\ref{derivativeKB}), it is followed with 
\beq
\frac{\partial K_B(t)}{\partial t}\approx 0, 
\eeq
for any large finite time. This directly proves the freezing behavior of $K_B(t)$. The analysis for the qubit entanglement $K_A(t)$ is similar.

\subsection{Freezing and thawing timings for finite $N$}
We first analyze the period of entanglement freezing and thawing alternation. Again, we focus on the time dependent amplitude $c_e(t)$ 
\begin{eqnarray}
c_{e}\left( t\right)  &=&\frac{2}{N+2}\sum_{k=1}^{N+1}\exp \left[ -iE_{k}t
\right] \sin ^{2}\left( \frac{k\pi }{N+2}\right), \label{ce}  \\
E_{k} &=&2\eta_{\mu \perp}\cos \left( \frac{k\pi }{N+2}\right). 
\end{eqnarray}
It is noted in the summation of $c_e(t)$ the terms when $k$ is close to $(N+2)/2$ make the largest contributions. Therefore, it is appropriate to expand the eigen energies by taking the first order approximation as 
\begin{eqnarray}
E_{k} &=&2\eta_{\mu \perp}\cos \left( \frac{k\pi }{N+2}\right) \simeq 2\eta_{\mu \perp}\left( \frac{\pi }{2}-\frac{k\pi }{N+2}\right) \notag \\
&&=\eta_{\mu \perp}\pi \left( 1-
\frac{2k}{N+2}\right),
\end{eqnarray}
which is an equally spaced spectrum. 

For such a spectrum, the revival time is inversely proportional to the common factor of all eigen energies (simply given by the neighboring energy space $\Delta E_k$) and can be obtained as
\begin{equation}
T=\frac{2\pi }{\Delta E_{k}}=\frac{2\pi }{2\eta_{\mu \perp}\pi }(N+2)=\frac{N+2}{\eta_{\mu \perp}}.
\end{equation}
This is exactly the period of the freezing-thawing alternation.

On the other hand, he time dependent coefficient can be rewritten as
\beqa
c_{e}( t) &\simeq& \sum_{k=1}^{N+1}\exp \left[  \frac{-i\pi\eta_{\mu \perp}t( N+2-2k)}{N+2}\right]\notag\\
&&\times \frac{2}{N+2}\sin ^{2}\left( \frac{k\pi }{N+2}\right) 
\eeqa
Entanglement freezing occurs when $c_{e}( t) =0$ corresponding to 
\beq
\eta_{\mu \perp}t=l(N+2)+2,
\eeq
where $l=0,1,2,3...$. Obviously, period information $T=(N+2)/\eta_{\mu \perp}$ is confirmed. Also the first freezing occurs at time $\eta_{\mu \perp}t=2$ independent of $N$.

\subsection{Freezing behavior independent of $N$}
In this section, we show why the behaviors at early times of entanglements $K_A(t)$, $K_B(t)$ for arbitrary finite $N$ are almost identical to the infinite $N$ case. We focus on the amplitude of the excited state as given in Eq.~(\ref{ce}) and perform a Taylor expansion with respect to $t$ to obtain 
\beqa
c_e(t) &=& \sum_{n=0}^{\infty} \frac{2^{n+1}}{N+2} \frac{1}{n!} (-i\eta_{\mu \perp}t)^n \notag \\
&&\bigg[\sum_{k=1}^{N+1}\cos^n\left(\frac{k\pi}{N+2}\right)\sin^2\left(\frac{k\pi}{N+2}\right)\bigg].
\eeqa
For $n=\text{odd}$, one has $c_e(t)=0$ because the term $\cos^n(\theta)\sin^2(\theta)$, with $\theta=k\pi/(N+2)\in(0, \pi)$, is asymmetric with respect to $\theta=\pi/2$. Only the even terms will survive and one can always define $n=2m$ with $m=0, 1, 2, \ldots$ to achieve 
\begin{equation}
c_e(t) = \sum_{m=0}^{\infty} \frac{2^{2m+1}}{N+2} \frac{1}{(2m)!} (-\eta_{\mu \perp}^2t^2)^{m} Y_m,
\end{equation}
where
\begin{equation}
Y_m =\sum_{k=1}^{N+1}\cos^{2m}(\theta) \sin^2(\theta).
\end{equation}

Then one can achieve the following new expression of the excited state amplitude
\begin{widetext}
\beqa \label{expansion}
c_e(t) &=& \sum_{m=0}^{\infty} (-\eta_{\mu \perp}^2t^2)^m \bigg[\underbrace{\frac{1}{m!(m+1)! }}_\text{for any $m$} 
+\underbrace{\frac{2(-7+m-8N-2N^2)}{\Gamma(m-N)\Gamma(m+N+4)}}_\text{when $m>N$} +\underbrace{\frac{2(-31+m-32N-8N^2)}{\Gamma(m-2N-2)\Gamma(m+2N+6)}}_\text{when $m>2N+2$}  + \ldots \bigg] \notag \\
&=&  \underbrace{c_e(t)}_\text{for $N\rightarrow \infty$} +  \sum_{m=N+1}^{\infty}(-\eta_{\mu \perp}^2t^2)^m \bigg[\underbrace{\frac{2(-7+m-8N-2N^2)}{\Gamma(m-N)\Gamma(m+N+4)}}_\text{when $m>N$} +\underbrace{\frac{2(-31+m-32N-8N^2)}{\Gamma(m-2N-2)\Gamma(m+2N+6)}}_\text{when $m>2N+2$}  + \ldots \bigg] 
\eeqa
\end{widetext}
where $\Gamma(n)=(n-1)!$ is the normal Gamma function. Here we have used the following series expansion identity 
\beqa
\cos^{2m}(\theta)& =& \frac{1}{2^{2m}}\left( 
\begin{array}{c}
2m \\
m
\end{array} 
\right) \notag \\
&&+
 \frac{2}{2^{2m}}\sum_{l=0}^{m-1}
\left( 
\begin{array}{c}
2m \\
l
\end{array} 
\right)
\cos\left[2(m-l)\theta\right].
\eeqa

One notes from (\ref{expansion}) that the first $N$ terms of the finite $N$ expansion is exactly the same as the infinite $N$ (i.e., Bessel-relted function $J_{1}(2\eta_{\mu \perp}t )/\eta_{\mu \perp}t$ as proved in Appendix A) case. This leads to the result that their behaviors are almost identical for $\eta_{\mu \perp}t < N+2$ due to the Taylor's theorem for estimates of the remainder. Therefore, one can conclude that for finite $N$ there is also freezing behavior. Combining with the proof the periodicity of revivals, this has confirmed the alternative appearances of freezing and thawing behaviors of the entanglements $K_A(t)$ and $K_B(t)$ shown in Fig.~\ref{freezing} (b) and (c) of the main text.


\begin{thebibliography}{99}





\bibitem{NC2000} M.A. Nielsen and I.L. Chuang, {\em Quantum Computation and Quantum Information} (Camb. Univ. Press, 2000).


\bibitem{Horodecki2009RMP} Ryszard Horodecki, Paweł Horodecki, Michał Horodecki, and Karol Horodecki, ``Quantum entanglement", \rmp{81}, 865 (2009).


\bibitem{Streltsov2017RMP} Alexander Streltsov, Gerardo Adesso, and Martin B. Plenio, ``Colloquium: Quantum coherence as a resource", \rmp{89}, 041003 (2017).


\bibitem{OlZur} H. Ollivier and W. H. Zurek, "Quantum Discord: A Measure of the Quantumness of Correlations", Phys. Rev. Lett. \textbf{88}, 017901 (2001).

\bibitem{HendVedr} L. Henderson and V. Vedral, ``Classical, quantum and total correlations", \jpa{34}, 6899 (2001).

\bibitem{Datta} A. Datta, A. Shaji, and C. M. Caves, ``Quantum Discord and the Power of One Qubit", Phys. Rev. Lett. \textbf{100}, 050502 (2008).

\bibitem{Mazzo} L. Mazzola, J. Piilo, and S. Maniscalco, ``Sudden Transition between Classical and Quantum Decoherence", Phys. Rev. Lett. \textbf{104}, 200401 (2010).


\bibitem{Maziero2009PRA} J. Maziero, L. C. \'Celeri, R. M. Serra, and V. Vedral, ``Classical and Quantum Correlations under Decoherence", \pra{80}, 044102 (2009).

\bibitem{Pinto} J. P. G. Pinto, G. Karpat, and F. F. Fanchini, ``Sudden change of quantum discord for a system of two qubits", Phys. Rev. A \textbf{88}, 034304 (2013).


\bibitem{Cianciar} M. Cianciaruso, T. R. Bromley, W. Roga, R. Lo Franco, and G. Adesso, ``Universal freezing of quantum correlations within the geometric approach", Scientific Reports \textbf{5}, 10177 (2015).


\bibitem{Silva}  I. A. Silva, D. Girolami, R. Auccaise, R. S. Sarthour, I. S. Oliveira, T. J. Bonagamba, E. R. deAzevedo, D. O. SoaresPinto, and G. Adesso, Phys. Rev. Lett. \textbf{110}, 140501 (2013).


\bibitem{Paula} F. M. Paula, et al., Phys. Rev. Lett. \textbf{111}, 250401 (2013).

\bibitem{AliRau} M. Ali and A. R. P. Rau, Phys. Rev. A \textbf{90}, 042330 (2014).

\bibitem{Bromley2015PRL} T.R. Bromley, M. Cianciaruso, and G. Adesso, \prl{114} 210401 (2015).

\bibitem{WuXu} W. Wu and J.-B. Xu, Journal of Physics B: Atomic, Molecular and Optical Physics \textbf{49}, 115502 (2016).

\bibitem{Ding} Y. Ding, S-B Xie and J.H. Eberly,  \pra{103}, 032418 (2021).


\bibitem{Kimble2008Nature} J.H. Kimble, ``The quantum internet", \nat{453}, 1023-1030 (2008).


\bibitem{Duan2003PRL} L.-M. Duan, E. Demler, and M. D. Lukin, ``Controlling Spin Exchange Interactions of Ultracold Atoms in Optical Lattices", \prl{91}, 090402 (2003).


\bibitem{Schrieffer-Wolff}  J. R. Schrieffer and P. A. Wolff, ``Relation between the Anderson and Kondo Hamiltonians", \pr{149}, 491 (1966).


\bibitem{Hewson1997} A.C. Hewson, ``The Kondo Problem to Heavy Fermions", (Cambridge University Press, Cambridge, England, 1997).


\bibitem{Qian2005PRA} See for an example in X.-F. Qian, Y. Li, Y. Li, Z. Song, and C. P. Sun, \pra{72}, 062329 (2005).


\bibitem{Grobe} R. Grobe, K. Rz\c{a}zewski and J. H. Eberly, \jpb {27}, L503 (1994).

\bibitem{Eberly2006} J. H. Eberly, Laser Phys. {\bf 16}, 921 (2006).


\bibitem{Qian2018NJP} A similar analysis can be found in X.-F. Qian, M.A. Alonso, J.H. Eberly, \njp{20} 063012 (2018).


\bibitem{PrattEberly} J.S. Pratt and J.H. Eberly, \prb {64}, 195314 (2001).

\end{thebibliography}
\end{document}